\documentclass[aps,showpacs,preprintnumbers,amsmath,amssymb,nofootinbib, twocolumn]{revtex4}
\usepackage{graphicx,amsmath,amssymb,color,multirow}

\renewcommand{\vec}[1]{\boldsymbol{\mathrm{#1}}}
\renewcommand{\Im}{{\rm Im~}}
\renewcommand{\Re}{{\rm Re~}}

\begin{document}

\title{Evolution and dynamical properties of Bose-Einstein condensate dark matter stars}
\author{Eniko J. M. Madarassy}
\email{eniko.madarassy@physics.uu.se}
\affiliation{Division of Astronomy and Space Physics, Uppsala University, 751 20 Uppsala, Sweden}
\author{Viktor T. Toth}
\email{vttoth@vttoth.com}
\affiliation{Ottawa, Ontario, Canada}

\begin{abstract}
Using recently developed nonrelativistic numerical simulation code, we investigate the stability properties of compact astrophysical objects that may be formed due to the Bose-Einstein condensation of dark matter. Once the temperature of a boson gas is less than the critical temperature, a Bose-Einstein condensation process can always take place during the cosmic history of the universe. Due to dark matter accretion, a Bose-Einstein condensed core can also be formed inside massive astrophysical objects such as neutron stars or white dwarfs, for example.

Numerically solving the Gross-Pitaevskii-Poisson system of coupled differential equations, we demonstrate, with longer simulation runs, that within the computational limits of the simulation the objects we investigate are stable. Physical properties of a self-gravitating Bose-Einstein condensate are examined both in non-rotating and rotating cases.

\end{abstract}

\maketitle

\section{Introduction}

A central issue in modern cosmology is the dark matter problem (see \cite{Sal2011} for an extensive review of recent results in the search for dark matter). Considering the existence of dark matter on galactic and extragalactic scales is necessitated by two fundamental sets of observations: the rotation curves of spiral galaxies and the mass discrepancy in clusters of galaxies.

Galaxy rotation curves \cite{BT1987, Per1996, Bor2001} compellingly indicate a failure of Newtonian gravity or standard general relativity on these scales in the absence of dark matter. The conflict between predicted and observed rotation curves can be resolved by postulating the existence of invisible dark matter forming a spherical halo around galaxies. Similarly, postulating dark matter can account for the apparent deficit in the virialized mass of galaxy clusters.

In most popular models, dark matter is cold and pressureless. A generic class of candidate dark matter is known by the acronym WIMP (Weakly Interacting Massive Particles---for a review of the particle physics aspects of dark matter, see \cite{Ov2004}.)

Candidate dark matter particles have extremely small but nonzero interaction cross sections with normal baryonic matter. Therefore, their direct experimental detection may be possible. Dark matter heating due to WIMP annihilation processes was also proposed as an alternative to fusion as the process that powered early, first-generation stars in the universe \cite{Spol2009,Freese2009,Freese2010}, resulting in a {\em dark star}. These models are predicated on the existence of antiparticle partners to proposed dark matter particles, but can potentially lead supermassive dark stars (SMDS) with masses of the order of $(10^5 - 10^7)M_{\odot}$ \cite{Freese2010}.

Inert or feebly annihilating dark matter, on the other hand, may play a role in present-day main sequence stars. Even a small spin-dependent interaction cross-section ($\sim 10^{-37}$~cm$^2$) between dark matter particles and protons can provide an energy transport mechanism with dramatic effects, limiting the mass of such dark matter particles to values greater than 5 GeV \cite{Iocco2012}.

Particles in a dilute Bose gas can form a quantum degenerate condensate (Bose-Einstein condensate, BEC) by occupying the same quantum state at low temperatures. Such states have been investigated in the laboratory using laser and evaporative cooling techniques (for recent reviews see \cite{Dal1999,Cor2002,Ket2002,Pit2003,Dui2004,Pethick2008}.)

For a particle of mass $m$ in an ensemble in thermodynamic equilibrium at temperature $T$, the thermal wavelength is
\begin{align}
\lambda _T =\sqrt{\frac{2\pi\hbar^2}{k_BmT}},
\end{align}
where $k_B$ is Boltzmann's constant. When the thermal wavelength exceeds the interparticle distance $l$, overlapping particles become correlated. The interparticle distance and the particle density $n$ are connected by the relationship $nl^3\simeq 1$. Therefore, the condition $\lambda_T > l$ can be re-expressed as $n\lambda_T^3 > 1$, yielding the transition temperature \cite{LL1976,LL1978}:
\begin{align}
T_c=\frac{2\pi\hbar^2n^{2/3}}{k_Bm}.
\end{align}

In the astrophysical context, BEC phenomena have been studied as a possible source of dark matter \cite{2007JCAP...06..025B}. Rotating BEC halos were also discussed \cite{Guzman2014}. The possibility that the superfluid interior of compact astrophysical objects may be at least partially in the form of a BEC \cite{Cha2012}, and the possible presence of a boson condensate in stars and their stability properties \cite{Harko2012} were also investigated. In these contexts, the BEC can be described as a non-relativistic degenerate gas with a barotropic equation of state.

However, the stability of self-gravitating BECs remains an open question. Chavanis and Harko \cite{Cha2012} studied the dynamics of a non-rotating BEC star and found it to be stable. More recently, this result was confirmed using a simple thermodynamic argument \cite{Lat2014}. On the other hand, Guzman \cite{Guzman2013} found that the self-gravitating BEC halos described in \cite{Cha2012} are unstable, and that furthermore, a relationship exists between the halo size and the BEC self-interaction parameter, which may be untenable (see also \cite{2011PhRvD..84d3531C}). However, Toth \cite{Toth2014a} showed that these difficulties arise from the inappropriate application of the Thomas-Fermi approximation, and recommended the use of a different density profile as an approximate static solution.

In the present paper, we investigate the stability of self-gravitating BECs using numerical software code that was originally developed to study two-dimensional condensates in the laboratory \cite{2008JLTP..152..122M,2009GApFD.103..269M,2009CaJPh..87.1013M,Madarassy2010}. The code was later extended to study three-dimensional self-gravitating condensates \cite{Madarassy2013} and further refined to eliminate instabilities due to the choice of the initial condensate profile \cite{Toth2014a}. The hypothetical objects of our study are made of ultralight bosons. This is dictated by computational necessity, not physics. The spatial resolution of the simulation must be below the Compton wavelength $\lambda_C$. Therefore, to keep the number of iterations computationally manageable, the linear scale of the simulation volume cannot be more than a few hundred times $\lambda_C$.

Nonetheless, even with this restriction in place, our approach can be employed to study a range of astrophysical objects using modest computational resources (e.g., desktop computers). Our investigations focus on compact BEC stars and extended BEC galactic halos. In this paper, we report our findings on the stability of BEC stars; results of our on-going efforts to study extended galactic halos will be reported elsewhere.

\section{Bose-Einstein condensates}

Below the critical temperature, $T<T_c$, bosons have a tendency to occupy a single quantum mechanical state, forming a Bose-Einstein condensate (BEC). This results in quantum effects becoming evident on a macroscopic scale.

\subsection{The Gross-Pitaevskii equation}
\label{sec:GPE}

If we treat a BEC as a classical object, the wave function of the condensate satisfies the non-linear Schr\"odinger equation (NLSE), also called the Gross-Pitaevskii equation (GPE) \cite{Dal1999,Cor2002,Ket2002,Pit2003,Dui2004,Pethick2008}. Therefore, the dynamics of weakly interacting particles of confined and dilute BECs below the critical temperature are described by a mean-field time-dependent macroscopic wavefunction, $\Psi(t,\vec{r})$, which we normalize such that
\begin{equation}
\int |\Psi|^2d^3 \vec{r}=N,
\label{eqn:Norm}
\end{equation}
where $N$ is the total particle number. The wavefunction is related to the condensate density $\rho(t,\vec{r})=m|\Psi(t,\vec{r})|^2$ (where $m$ is the particle mass) and phase, $\varphi=\tan^{-1}(\Im\Psi/\Re\Psi)$ through the Madelung transformation, $\Psi=|\Psi|e^{i\varphi}=\sqrt{m^{-1}\rho}e^{i\varphi}$.

Inter-particle collisions are described by the s-wave scattering length, denoted by the letter $a$. In a sufficiently dilute condensate, characterized by an average interparticle distance that is much greater than $a$, particle interactions are binary interactions that are characterized by a contact interaction potential, $V_{\rm int}(\vec{r},\vec{r}')=g\delta (\vec{r}-\vec{r}')$, where $g$ is the scattering coefficient and $\delta$ is the Dirac $\delta$-function. The coupling constant $g$ is related to the scattering length by the equation
\begin{equation}
g=\frac{4\pi\hbar^{2}a}{m}.
\label{eqn:coupl_const1}
\end{equation}
This effective interaction is attractive if $g<0$, repulsive otherwise. The $N$-particle Hamiltonian that corresponds to this pseudopotential can be written as
\begin{equation}
\hat{H}=\sum\limits_{i=1}^N\left[-\frac{\hbar^2}{2m}\nabla_i^2+V(\vec{r}_i)\right]+\sum\limits_{i<j}g\delta(\vec{r}_i-\vec{r}_j),
\end{equation}
where $\nabla_i=\partial/\partial\vec{r}_i$ and $V(\vec{r}_i)$ represents external potentials. This leads to the following non-linear Schr\"odinger-type equation (NLSE):
\begin{equation}
i\hbar\frac{\partial\Psi}{\partial t}=\left[-\frac{\hbar^{2}}{2m}\nabla^{2}+V-\mu+g|\Psi|^{2}\right]\Psi,
\label{eqn:GPE}
\end{equation}
which is the Gross-Pitaevskii equation \cite{Dal1999,Cor2002,Ket2002,Pit2003,Dui2004,Pethick2008}.

In (\ref{eqn:GPE}), we also introduced the chemical potential $\mu$. In a thermodynamical system, the chemical potential is defined as the amount $\mu=\partial E/\partial N$ by which the internal energy $E$ of the system changes if we introduce an additional particle while keeping the entropy and volume fixed. Below the critical temperature, $T<T_c$, the chemical potential is well approximated by $\mu(T)=-kT\ln(1+1/N)\simeq 0$, and thus it can be ignored for large $N$. Hence from now on, we assume $\mu=0$.

\subsection{Self-gravitating systems}

In a self-gravitating system with no external potential, the potential $V$ is the gravitational potential that is determined from Poisson's equation for gravity:
\begin{equation}
\nabla^2V=4\pi G\rho=4\pi Gm|\Psi|^2,
\end{equation}
where $G$ is Newton's gravitational constant. Together with Eq.~(\ref{eqn:GPE}), these equations form what is known the Gross-Pitaevskii-Poisson (GPP) system of equations.

For a self-gravitating system, the GPE can be derived using the variational principle from the energy functional (cf. \cite{Wang2001,Pethick2008}):
\begin{align}
{\cal E}&=\frac{\hbar^2}{2m}|\nabla\Psi|^2+\frac{1}{2}V|\Psi|^2+\frac{1}{2}g|\Psi|^4.
\end{align}
(Note that the factor of $1/2$ multiplying $V$ is required to avoid double counting the gravitational potential energy between two regions of the condensate.)

From (\ref{eq:E}), the total energy of the system can be calculated as
\begin{equation}
E_{\rm tot}=E_{\rm kin}+E_{\rm pot}+E_{\rm int},\label{eq:E}
\end{equation}
where the kinetic energy $E_{\rm kin}$, (gravitational) potential energy $E_{\rm pot}$ and internal energy $E_{\rm int}$ are given by, respectively:
\begin{align}
E_{\rm kin}&=\int\frac{\hbar^2}{2m}|\nabla\Psi|^2~d^3\vec{r},\\
E_{\rm pot}&=\int\frac{1}{2}V|\nabla\Psi|^2~d^3\vec{r},\\
E_{\rm int}&=\int\frac{1}{2}g|\Psi|^4~d^3\vec{r}.
\end{align}

To achieve a stable self-gravitating object, it is necessary to choose a coupling coefficient (\ref{eqn:coupl_const1}) that is positive. For an initial estimate $a_0$ for the scattering length $a$, we use the value
\begin{equation}
a_0=Gm^3(R/\pi\hbar)^2,
\end{equation}
which is derived from the Thomas-Fermi approximation \cite{2007JCAP...06..025B,2011PhRvD..84d3531C}. Since we do not rely on that approximation, other values for $a$ are of course also possible.

\subsection{Quantization of circulation}

In a frame of reference that is rotating in the $xy$-plane with angular velocity $\Omega$, the GPE is written as
\begin{equation}
\label{eqn:GPE1}
i\hbar\frac{\partial\Psi}{\partial t} =\left[-\frac{\hbar^{2}}{2m}\nabla^{2}+V+g|\Psi|^{2}-\Omega L_z\right]\Psi,
\end{equation}
where $L_z=-i\hbar(x\partial_{y}-y\partial_{x})$.

Formation of observable of quantum vortices represents one the most remarkable properties of BECs. The GPE permits topologically non-trivial solutions, such as vortices with non-zero fluid circulation and zero density. These vortices appear as density holes with quantized circulation. Vortices may decay by colliding, or dissipate at the condensate surface or through other dissipative mechanisms. Vortex creation is influenced by the shape and size of the condensate and the nature and form of the potential $V$.

When a superfluid rotates about a fixed axis at a low enough frequency, it remains stationary. When the rotation frequency exceeds a critical value, quantized vortex lines may appear, vortices form near the surface \cite{Bay1996} and enter the condensate. The emergence of vortices reduces the free energy of the system and they become energetically favorable \cite{Fet2001}.

The GPE is formally similar to the Ginzburg-Landau equation, which describes a superfluid with $\Psi$ as a complex order parameter field. Any rotation of the fluid must be in the form of vortex lines, which introduce quantum circulation. The phase $\varphi$ of the order parameter around any closed contour $K$ must be $2\pi q$, where $q=0,\pm1,\pm2,...$:
\begin{equation}
\oint_K \nabla\varphi\cdot d\vec{l}=2\pi q.
\label{eqn:PACC}
\end{equation}
The gradient of the phase describes the local velocity flow: $\vec{v}=\hbar m^{-1}\nabla\varphi$ \cite{Pethick2008}. The superfluid rotation induced by a vortex line can be expressed as the circulation, $\Gamma$, about $K$:
\begin{equation}
\Gamma = \oint_K \vec{v}\cdot d\vec{l}=2\pi q\frac{\hbar}{m},
\label{eqn:circulation}
\end{equation}
which is an integer multiple of $2\pi\hbar/m$. This shows that circulation is quantized in units of $2\pi\hbar/m$. For $|q| > 1$ the system is unstable. Also from Eq.~\eqref{eqn:circulation} it follows that the velocity around a single circular vortex of radius $r$ is:
\begin{equation}
v=\frac{\Gamma}{2\pi r}=\frac{q\hbar}{mr}.
\end{equation}
There is an energy barrier between non-vortex and vortex states. The superfluid has identically vanishing vorticity ($\nabla\times\vec{v}=\hbar m^{-1}\nabla\times\nabla\varphi = 0$) everywhere excluding singular (vortex) lines. The minimum density grows to the bulk value over a length-scale of order the healing length \citep{Bay1996}. As we move away from the vortex the velocity slowly decreases. If we move towards the vortices then the superfluid density tends to zero.

\subsection{Choice of units}
\label{sec:units}

In the remainder of this paper, we choose to use a dimensionless form of the GPE, which is obtained by using units such that $\hbar=1$ and $m=1$:
\begin{equation}
\label{eqn:GPE2}
i\frac{\partial\Psi}{\partial t}=\left[-\frac{1}{2}\nabla^{2}+V+g|\Psi|^{2}-\Omega L_{z}\right]\Psi.
\end{equation}
This choice fixes two of the three fundamental units (e.g., time and mass), whereas the third unit can be chosen as a matter of convenience. After restoring units, the total condensate mass can be obtained as $M=Nm$.

We note that the dimensionless GPP system is invariant under the following set of transformations:
\begin{align}
t \rightarrow \lambda^{2} t,\qquad r &\rightarrow \lambda r,\qquad V \rightarrow \lambda^{-2} V,\nonumber\\
g \rightarrow \lambda^{-2} \kappa^{-2} g,\qquad G &\rightarrow \lambda^{-4} \kappa^{-2} G,\qquad \Psi \rightarrow \kappa \Psi.
\label{eqn:transformations}
\end{align}
Under these rescalings, the GPP system will read as
\begin{equation}
\frac{\kappa i}{\lambda^{2}}\frac{\partial \Psi}{\partial t} = \left[-\frac{\kappa}{2\lambda^{2}} \nabla^{2}
+ \frac{\kappa}{\lambda^{2}} V+ \frac{\kappa g}{\lambda^{2}}|\Psi|^{2}-\frac{\kappa\Omega}{\lambda^{2}} L_{z}\right]\Psi,
\label{eqn:RSGPPE}
\end{equation}
\begin{equation}
\lambda^{-4}\nabla^{2}V=\lambda^{-4} 4\pi G |\Psi|^2.
\label{eqn:RSPois}
\end{equation}
In particular, choosing a value of $\kappa\ne 1$ amounts to rescaling $N \rightarrow \kappa^2N$ and $G \rightarrow \kappa^{-2}G$. This rescaling is computationally convenient, as it allows all quantities to be represented as single-precision floating point numbers.

\section{The numerical code}

Our present study is based on a previously developed numerical solution of the GPE \cite{2008JLTP..152..122M,2009GApFD.103..269M,2009CaJPh..87.1013M,Madarassy2010}, which was obtained using the {\em Crank-Nicholson method} in combination with {\em Cayley's formula} \cite{NUMRC}, in the presence of an isotropic trapping potential (for a numerical investigation of BECs in the presence of anisotropic traps see \cite{Muru2009,Muru2012}.) In particular, the use of {Cayley's formula ensures that the numerical solution remains stable, and the unitarity of the wavefunction is maintained.

The code was later extended to study three-dimensional self-gravitating condensates \cite{Madarassy2013} by incorporating a solution of the gravitational Poisson equation using the {\em relaxation method}, and by introducing a novel initial estimate for the condensate density to eliminate instabilities \cite{Toth2014a}.

\subsection{Discretization of the GPE}

Given the time-dependent Schr\"odinger's equation in the general form,
\begin{equation}
i \frac{\partial\Psi}{\partial t}=\hat{H}\Psi,\label{eq:GPE1}
\end{equation}
the value $\Psi^{n+1}$ of the wavefunction at the $(n+1)$-th time step is obtained from the known values $\Psi^n$ at the $n$-th time step by solving
\begin{equation}
\left(1+\frac{1}{2}i\Delta t\hat{H}\right)\Psi^{n+1}=\left(1-\frac{1}{2}i\Delta t\hat{H}\right)\Psi^n.\label{eq:CrankNich}
\end{equation}
After evaluating the right-hand side given $\Psi^n$, the left-hand side can be solved for $\Psi^{n+1}$ algebraically. If $\hat H$ is a linear operator, this is a linear system of equations for the unknown values $\Psi^{n+1}$.

This is elegantly demonstrated in the one-dimensional case, where the Hamilton operator is given by
\begin{equation}
\hat{H}=-\frac{1}{2}\frac{\partial^2}{\partial x^2}+U.
\end{equation}
The second derivative can be approximated as a finite difference as follows:
\begin{equation}
\frac{\partial^2\Psi}{\partial x^2}=\frac{\Psi_{k-1}-2\Psi_k+\Psi_{k+1}}{(\Delta x)^2},
\end{equation}
Substituting the above expression into Eq.~(\ref{eq:CrankNich}), we find
\begin{align}
&\left[1-\frac{i\Delta t}{2}\left(U+\frac{1}{(\Delta x)^2}\right)\right]\Psi^{n+1}_k - \frac{i\Delta t}{4(\Delta x)^2}\left(\Psi^{n+1}_{k-1}+\Psi^{n+1}_{k+1}\right)\nonumber\\
&=\left[1-\frac{i\Delta t}{2}\left(U+\frac{1}{(\Delta x)^2}\right)\right]\Psi^n_k+ \frac{i\Delta t}{4(\Delta x)^2}\left(\Psi^n_{k-1}+\Psi^n_{k+1}\right).
\label{eq:tridiag}
\end{align}
If the values of $\Psi^n$ on the right-hand side are known, Eq.~(\ref{eq:tridiag}) represents  an algebraic system of equations for the values of $\Psi^{n+1}$ that appear on the left-hand side. The equations are linear if the generic potential $U$ is not itself a function of $\Psi$. Moreover, the form of this system of equations is tridiagonal, and therefore it can be solved highly efficiently by using the {\em Thomas algorithm} \cite{NUMRC}.

In the three-dimensional case, one could proceed with solving for $\Psi^{n+1}$ directly, but since the system (\ref{eq:tridiag}) is no longer tridiagonal, the efficiency of the numerical procedure is lost. Therefore it is better to use the {\em alternating-direction implicit method}, which requires the calculation of the one-dimensional solution in the $x$, $y$ and $z$ directions, using $\Delta t/3$ for the time step, and $U/3$ for the potential. This method works  because the Hamilton operator can be represented as a sum of three operators, $\hat{H}=\sum _{i=1}^3\hat{H}_i$, $i=x,y,z$, allowing us to solve numerically using fractional time steps as follows:
\begin{equation}
\hat{H}_i=-\frac{1}{2}\frac{\partial^2}{\partial x_i^2}+\frac{1}{3}U,\qquad {\rm for}~i=x,y,z.
\end{equation}

Substituting these expressions into Eq.~(\ref{eq:tridiag}), that is, replacing $U$ with $U/3$ and $\Delta x$, respectively, with $\Delta x$, $\Delta y$ or $\Delta z$, and by using a time step of $\Delta t/3$, we obtain the three fractional iteration steps that correspond to a full iteration with time step $\Delta t$.

In the case of the GPE, the generic potential $U$ is non-linear, as it includes the self-interaction potential $g|\Psi|^2$. This non-linear term can be dealt with by a simple recursion that converges rapidly. Specifically, we calculate the non-linear term on the left-hand side by substituting $\Psi^n$ in place of $\Psi^{n+1}$, and solve the system of equations; we then use this solution to recalculate the non-linear term and solve again, until convergence is obtained. As the non-linear term is cubic, convergence is rapid.

\subsection{Numerical solution of Poisson's equation}

For the solution of Poisson's equation, we employ the {\em relaxation method}, a moderately efficient algorithm. This method is based on the finite differences approximation of the second derivative in Poisson's equation:
\begin{align}\label{Poi}
\nabla^2 V=&\frac{\partial^2V}{\partial x^2}+\frac{\partial^2V}{\partial y^2}+\frac{\partial^2V}{\partial z^2}\nonumber\\
=&{}\phantom{+~}\frac{V(x-\Delta x,y,z)-2V(x,y,z)+V(x+\Delta x,y,z)}{\Delta x^2}\nonumber\\
&{}+\frac{V(x,y-\Delta y,z)-2V(x,y,z)+V(x,y+\Delta y,z)}{\Delta y^2}\nonumber\\
&{}+\frac{V(x,y,z-\Delta z)-2V(x,y,z)+V(x,y,z+\Delta z)}{\Delta z^2}\nonumber\\
&{}+{\cal O}(\Delta^2),
\end{align}
where $\Delta^2$ is the largest of $\Delta x^2$, $\Delta y^2$ and $\Delta z^2$.

Equation~(\ref{Poi}) can be solved for $V(x,y,z)$. The relaxation method amounts to obtaining refined approximations of $V(x,y,z)$ using the iteration formula
\begin{align}
V_{k+1}&(x,y,z)=\frac{\Delta x^2\Delta y^2\Delta z^2}{2(\Delta x^2\Delta y^2+\Delta y^2\Delta z^2+\Delta z^2\Delta x^2)}\\
{}\times\bigg[&\phantom{+~}\frac{V_k(x-\Delta x,y,z)+V_k(x+\Delta x,y,z)}{\Delta x^2}\nonumber\\
&{}+\frac{V_k(x,y-\Delta y,z)+V_k(x,y+\Delta y,z)}{\Delta y^2}\nonumber\\
&{}+\frac{V_k(x,y,z-\Delta z)+V_k(x,y,z+\Delta z)}{\Delta z^2}-4\pi G|\psi|^2\bigg].\nonumber
\end{align}

This method is accurate, but does not converge rapidly. Its efficiency in the context of the GPP system can be improved, however, by observing that $|\psi|^2$ changes very little between subsequent timesteps. therefore, using the values of $V(x,y,z)$ from the preceding timestep as the inital estimage for $V(x,y,z)$ in the new iteration leads to very rapid convergence, with a satisfactory solution often emerging after just a few iterations.

For a full description and implementation of the numerical code see \cite{Madarassy2013}.

\subsection{Initial and boundary conditions}
\label{sec:INIT}

Numerically solving a system of coupled differential equations requires a set of initial and boundary conditions.
Specifically, solving the Gross-Pitaevskii equation requires an initial field configuration $\Psi(x,y,z)$ at $t=0$. In turn, the numerical solution of Poisson's equation for gravity needs boundary conditions in the form of values $V(x_{\rm min},y,z)$, $V(x_{\rm max},y,z)$, $V(x,y_{\rm min},z)$, $V(x,y_{\rm max},z)$, $V(x,y,z_{\rm min})$ and $V(x,y,z_{\rm max})$, respectively. We obtain suitable initial conditions by postulating an initial density profile in the form:
\begin{equation}
|\Psi|\propto \left|\frac{\sin r/r_0}{r/r_0}\right|^3,
\end{equation}
which is a good approximation of the numerical solution of the time-independent GPE \cite{Toth2014a}. We choose a value for $r_0$ to yield a condensate of the desired size, and a proportionality factor to obtain the desired total condensate mass. The phase of the condensate is initially vanishing: $\varphi=0$. As demonstrated in \cite{Toth2014a} and confirmed using the numerical code, this initial density profile yields stable numerical solutions.

During the simulation, the condensate is assumed to vanish, $\Psi=0$, at the boundaries of the simulation volume. The gravitational potential $V$ at the simulation volume is given by $V=-GN/r$ (where $r$ is the distance from the center of the simulation volume), corresponding to the Newtonian potential of a spherically symmetric condensate of $N$ particles.

\section{Model and results}

Presently, our research focused on the properties of a nonrotating or slowly rotating pure Bose star composed of ultralight bosons. (Stars that contain a mixture of bosons and ordinary matter require a hydrodynamical extension of our simulation code, which is not presently available.) The parameters of the Bose stars that we investigated are dictated both by physics and by the limits of our computational approach.

\subsection{The Bose star}

A Bose star is simulated with a mass $M=M_\odot=2\times10^{30}$~kg (one solar mass). The Schwarzschild radius corresponding to one solar mass is $R_S=2GM/c^2\simeq 3$~km ($c$ is the speed of light.) We chose a radius of $R=50$~km for our investigations, which is sufficiently large such that relativistic effects due to strong gravitational fields will remain small (of order $2GM/c^2R\simeq 0.06$), and the nonrelativistic GPE remains a valid approximation for a qualitative study.

The simulation volume must be significantly larger than the simulated star, to ensure that the star does not interact too much with the nonphysical boundaries of the volume. After experimentation, we chose a simulation box with coordinates that extend from $-440$ to $+440$ km.
Our simulation code divides space into cells, with the execution time being a polynomial function of the number of cells. Too many cells slow down the simulation unacceptably. For practical simulations, we use 80 cells, so the cell size is $11\times11\times11$ km.

If we take this cell size to be the Compton half-wavelength of the BEC particle, $\textstyle\frac{1}{2}\lambda=11$~km, the corresponding mass is $m=h/c \lambda\simeq10^{-46}~{\rm kg}\simeq 5.6\times10^{-11}$~eV. This is the maximum Bose particle mass that we can simulate using a spatial resolution of 11~km. We note that this mass is also consistent with the relativistic upper limit of $\sim 1.5\times 10^{-46}$~kg for a boson mass forming a $1~M_\odot$ star \cite{2012LRR....15....6L}, as well as some dark matter models involving ultralight bosons.

The time that corresponds to a Compton wavelength of 22~km is approximately 70~$\mu$s. This determines the maximum allowable simulation timestep. In most simulations, we shall use $\tau=10~\mu$s as our simulation timestep.

With these considerations in mind, we chose units for the purpose of simulating a BEC stellar remnant or stellar core as follows. First, we measure masses and lengths in the following units:
\begin{align}
1~[{\rm M}]&=10^{-46}~{\rm kg}\simeq 5.61\times10^{-11}~{\rm eV},\\
1~[{\rm L}]&=1~{\rm km}.
\end{align}
Given $\hbar\simeq 10^{-34}$~m$^2$kg~s$^{-1}$, these choices determine our unit of time:
\begin{equation}
1~[{\rm T}] = 1 \mu{\rm s}.
\end{equation}
In these units, $G=6.67\times10^{-78}~[{\rm L}]^3/[{\rm M}][{\rm T}]^2$. Furthermore, a Bose star of one solar mass, $M=2\times10^{30}$~kg, contains $N=2\times10^{76}$ particles.

These quantities can be rescaled (see subsection \ref{sec:units} above): Using $\kappa=10^{-38}$ allows us to rescale $G\rightarrow 0.0667$ and $N \rightarrow 2$. It must be emphasized that the physical conclusions are unaffected by this rescaling.

\subsection{Numerical stability}

As we experimented with our simulation code, there were indications that during a longer simulation run, the object may become unstable. {\em This behavior was expected}, due to the relatively coarse spatial and time resolution of the simulation, as well as the fact that the condensate unavoidably interacts with the simulation wall, where the boundary condition $|\Psi|=\phi=0$ is enforced. However, it was important to characterize the onset of any instability due to non-physical parameters of the simulation, and to distinguish these non-physical instabilities from real physical effects. Specifically, we investigated the impact of the finite size of the simulation volume element, the finite timestep, and the finite extent of the simulation volume.

In particular, we performed four long-duration, 100-second simulation runs (with runtimes amounting to several weeks each on desktop-class hardware) to see the onset and evolution of any computational instability. The parameters in these simulations were identical, except for the non-physical values of the time step and spatial resolution, which are summarized in Table~\ref{tab:runs1}.

\begin{table}
\caption{Four Bose star simulation runs with varying non-physical parameters.\label{tab:runs1}}
\begin{center}
\begin{tabular}{l|c|c}\hline\hline
~~Run&~~Time step~~&Spatial grid~~\\ \hline\hline
~~Case 1~~&10~$\mu$s&~~$80\times 80\times 80$~~\\ \hline
~~Case 2&20~$\mu$s& $~~80\times 80\times 80$~~\\ \hline
~~Case 3&\phantom{0}5~$\mu$s&~~$80\times 80\times 80$~~\\ \hline
~~Case 4&10~$\mu$s&$~~60\times 60\times 60$~~\\ \hline\hline
\end{tabular}
\end{center}
\end{table}

\begin{figure}
\includegraphics[width=\linewidth]{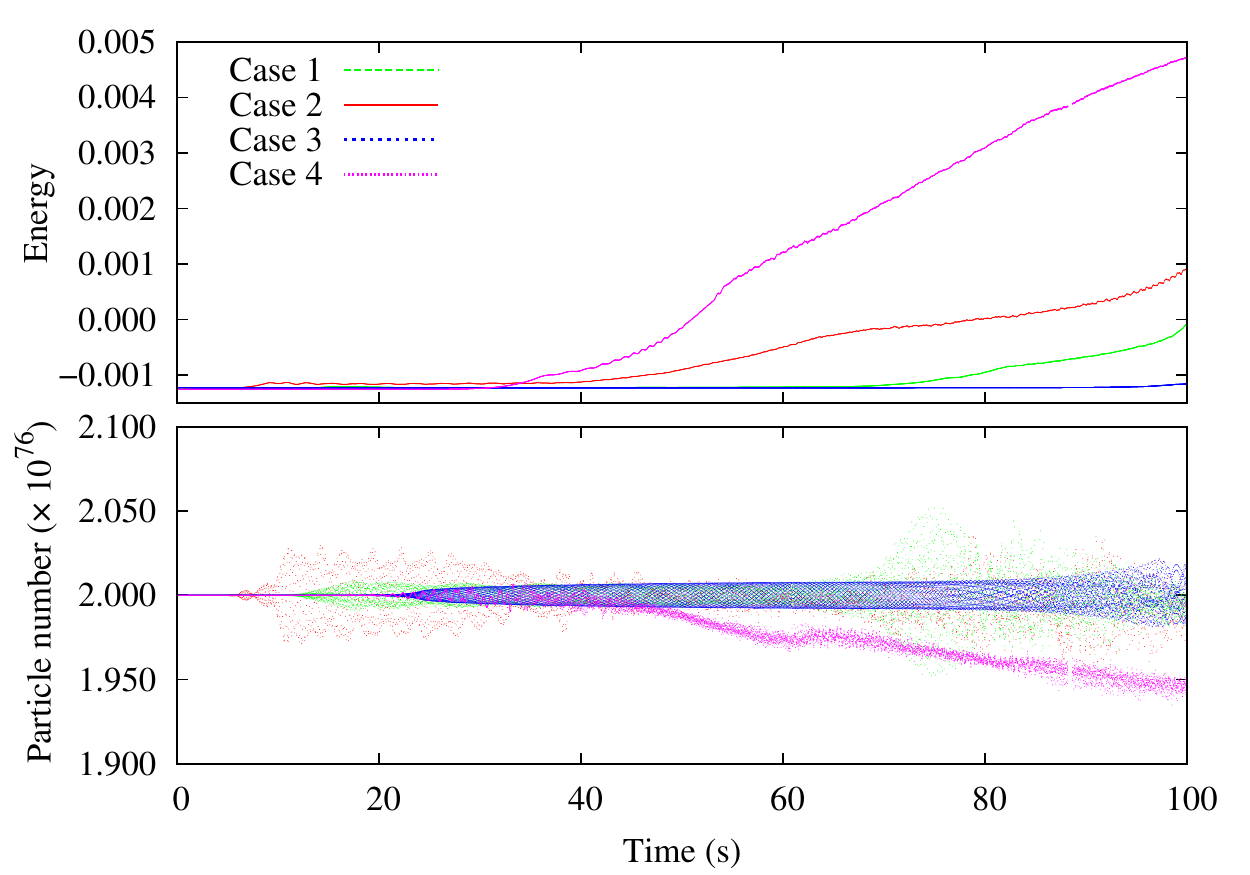}
\caption{\label{fig:EN}Non-conservation of energy and particle number as a result of the simulation interacting with nonphysical parameters.}
\end{figure}

The results are shown in Figure~\ref{fig:EN}. The stability of the simulation is characterized by observing supposedly conserved quantities, namely the total energy and particle number, in each of the four cases.

The principle of energy conservation dictates that the total energy should be constant. If it isn't, absent any programming errors, the result is due either to interaction with the walls of the simulation volume or numerical limitations. In the latter case, the outcome should depend on the choice of time step and volume element size in a predictable manner: the finer-grained the simulation is, the less the total energy should deviate from a constant value. And this is indeed what we see in the top panel of Fig.~\ref{fig:EN}.

The behavior of the total particle number is similar. The evolution of the GPE should be unitary; we did not introduce dissipation or other interactions with the environment that would change $N$. On the other hand, once again the finite size of the simulation volume and finite integration steps can affect the stability of $N$. As Fig.~\ref{fig:EN} shows, all simulations show a small oscillation appearing after approximately 10 seconds of simulation time. This oscillation may be a result of interaction with the walls. However, the magnitude of the oscillations, which are of $\mathcal{O}(1\%)$ and the point of onset of a systemic deviation from the conserved particle number both depend on the non-physical integration parameters. Videos of these simulations are available online \cite{bosestar-url}.

These results allowed us to find a conservative set of nonphysical parameters for simulations of ${\cal O}(1)$ seconds (corresponding to $\sim$160 low altitude orbits) of a compact Bose star, ensuring that the nonphysical parameters have no impact on the simulation outcome.

\subsection{Stability of stationary and rotating Bose stars}

Is a Bose star stable? Guzm\'an et al. argued \cite{Guzman2013} that it is not. However, the mass profiles \cite{2007JCAP...06..025B} that they examined were constructed on the basis of the Thomas-Fermi approximation. We found that the Thomas-Fermi profile leads to unstable numerical solutions; the reason for this was explored in Ref.~\cite{Toth2014a}.

The nonrotating Bose star that was the subject of our investigation of numerical stability described in the preceding section was found to be robustly stable within the numerical limits of the simulation code, but at least for several seconds. (While this may sound like a short amount of time for an astrophysical object, we are reminded that for a compact object of one solar mass and 50 km radius, the orbital period near the object's surface is only about 6~ms; therefore, a 1-second simulation run amounts to over 160 such orbits, which is more than sufficient to judge the stability of a self-gravitating object.)

Stability, in this case, was characterized by the fact that so long as the simulation remained numerically stable, the spatial distribution of the density $\rho$ remained constant, only the phase $\phi$ of the wavefunction $\Psi$ evolved as a function of time. From this we conclude that a stable Bose star is possible and consistent with the properties of the GPP system.

\begin{table}
\caption{The rotating Bose star.\label{tab:runs2}}
\begin{center}
\begin{tabular}{l|c|c}\hline\hline
~~Run&~~Box size~(km)~~& $\Omega$ (rad/s) \\ \hline\hline
~~Case 5~~&880& $\phantom{3\times{}}10^{-3}$ \\ \hline
~~Case 6&880& $3\times 10^{-4}$ \\ \hline
~~Case 7&880&~~no rotation~~\\ \hline
~~Case 8&280& $\phantom{3\times{}}10^{-4}$ \\ \hline
\hline
\end{tabular}
\end{center}
\end{table}

We also investigated a few cases (see Table~\ref{tab:runs2}) where the star was given an initial rotation, by introducing a nonzero $\Omega$ for an initial number of iterations. Given our choice of units, angular velocity is measured in radians per microsecond. For an $R=50$ km sphere, the equatorial circumference is $2\pi R \simeq 314$ km. A tangential velocity of $c=3\times 10^5$~km/s corresponds to $~955$ revolutions per second, or $\sim 0.006$~rad/$\mu$s. In order to remain within the approximately non-relativistic regime, therefore, angular velocities must be less than $10^{-3}$~rad/$\mu$s. Moreover, the orbital velocity of a one solar mass object at a radius of 50 km is $\sim 51.6\times 10^3$~m/s, which corresponds to $10^{-3}$~rad/$\mu$s.

We performed a test run of a very rapidly rotating condensate ($\Omega=10^{-3}$~rad/s). The condensate was unstable (see animations at \cite{bosestar-url}), which is hardly surprising, since the equatorial velocity was approximately equal to the orbital velocity. Interaction between the condensate and asymmetries in the simulation algorithm (e.g., the evaluation order singles out a preferred direction) resulted in an increasing ``wobble'', with the condensate ultimately colliding with the walls of the simulation volume. Clearly by this point, the simulation was unphysical.

When we used lower rotational velocities, however, the condensate remained stable over the simulated duration. These simulations, at angular velocities of $\Omega=3\times 10^{-3}$~rad/$\mu$s and  $\Omega=10^{-4}$~rad/$\mu$s corresponded to $\sim 53$ and $\sim 16$ full rotations, respectively, over the simulated duration of 1 second. Therefore, we conclude that our simulations demonstrate that even a rapidly rotating Bose star can remain stable.

\begin{table}
\caption{\label{tb:cases}The 36 cases in the parameter space that were investigated. See text for details.}
\begin{tabular}{cc|c|c|c|c|c|c}\hline\hline
\multicolumn{2}{r|}{$N=~$}&\multicolumn{2}{|c|}{$2\times 10^{76}$}&\multicolumn{2}{|c|}{$1\times 10^{76}$}&\multicolumn{2}{|c}{$4\times 10^{76}$}\\
$a/a_0$&$\Omega$\textbackslash$R=~$&50 km&80 km&50 km&80 km&50 km&80 km\\\hline\hline
\multirow{2}{*}{1}&0&{\scriptsize case 11}&{\scriptsize case 12}&{\scriptsize case 13}&{\scriptsize case 14}&{\scriptsize case 15}&{\scriptsize case 16}\\
&0.0001&{\scriptsize case 21}&{\scriptsize case 22}&{\scriptsize case 23}&{\scriptsize case 24}&{\scriptsize case 25}&{\scriptsize case 26}\\\hline
\multirow{2}{*}{0.5}&0&{\scriptsize case 31}&{\scriptsize case 32}&{\scriptsize case 33}&{\scriptsize case 34}&{\scriptsize case 35}&{\scriptsize case 36}\\
&0.0001&{\scriptsize case 41}&{\scriptsize case 42}&{\scriptsize case 43}&{\scriptsize case 44}&{\scriptsize case 45}&{\scriptsize case 46}\\\hline
\multirow{2}{*}{2}&0&{\scriptsize case 51}&{\scriptsize case 52}&{\scriptsize case 53}&{\scriptsize case 54}&{\scriptsize case 55}&{\scriptsize case 56}\\
&0.0001&{\scriptsize case 61}&{\scriptsize case 62}&{\scriptsize case 63}&{\scriptsize case 64}&{\scriptsize case 65}&{\scriptsize case 66}\\\hline\hline
\end{tabular}
\end{table}

\begin{figure}
\centering\includegraphics[width=0.95\linewidth]{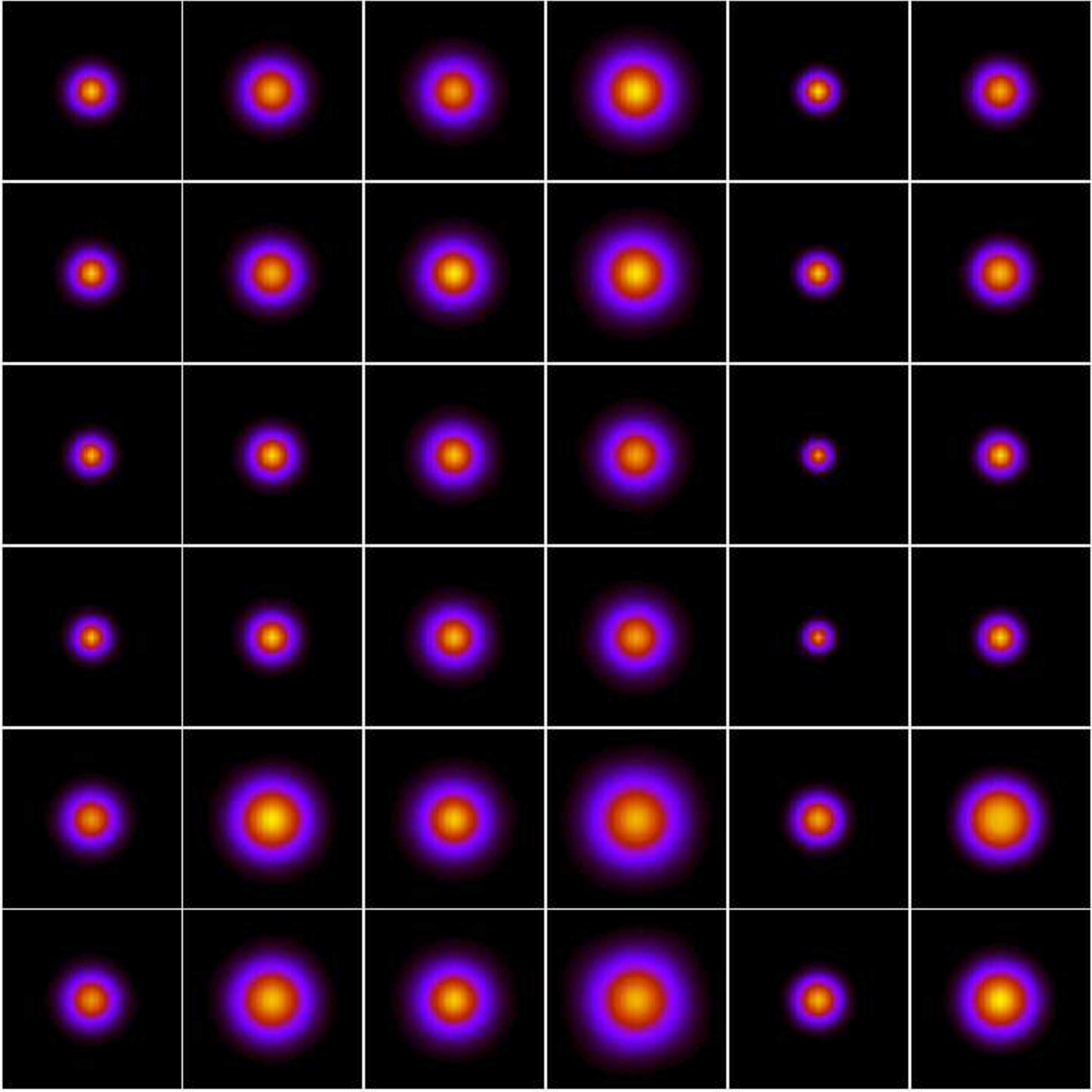}
\caption{\label{fig:allcases}End stages of the 36 simulations shown in the same order as in Table~\ref{tb:cases}, after 100,000 iterations corresponding to 1 second of elapsed time. Plots depict density cross-sections with an area of $480\times 480$~km$^2$ in the $xy$-plane at $z=0$, with $z$ being the rotation axis in a coordinate system centered on the Bose star. Cross-sections remained static during the simulations. Color scale is dynamic, for maximum contrast.}
\end{figure}

We were puzzled by the lack of any obvious formation of vortices. Viewing plots of the phase $\phi$, however, revealed possible vortex formation outside the nominal radius of the Bose star. This suggests that if any quantized vortices form as a result of rotation, these will form at the surface of the Bose star, and may not be able to penetrate the interior which has increasing density.

\subsection{Exploring the parameter space}

The parameter space of this simulation is vast, even if we leave nonphysical parameters out of consideration.

We began to explore systematically the parameter space by varying the values of the radius $R$ of the condensate, the total particle number (hence, the mass), cases of rotation vs. no rotation, and the magnitude of the coupling constant. Specifically, we investigated 36 different scenarios that are depicted in Table~\ref{tb:cases}, by varying $N$, $R$, $g$ and $\Omega$, this time in a $480\times 480\times 480$~km$^3$ simulation volume.

These 36 cases all yielded stable Bose stars (see Fig.~\ref{fig:allcases}, depicting the simulated objects after 100,000 iterations amounting to 1 second of simulation time.)

While the density plots of these Bose objects are essentially static, the phase of the wavefunction is strongly time dependent. Animations of these 36 cases are available online \cite{bosestar-url}. The phase diagrams also confirm what we learned earlier: that no vortex lines enter the interior of the object and if any vortices form, they do so in the very dilute regions at the object's surface.

\begin{figure}
\centering\includegraphics[width=\linewidth]{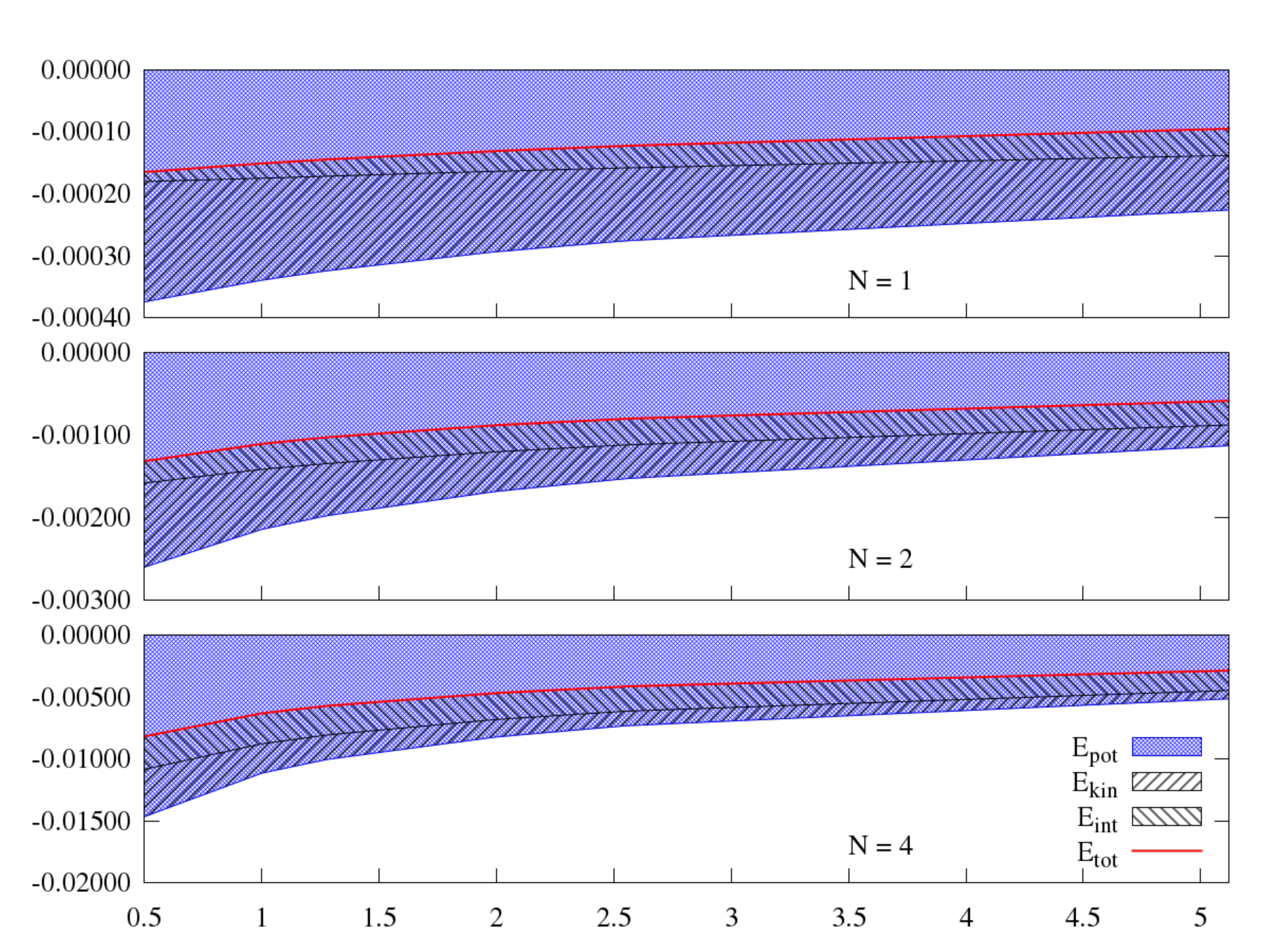}
\caption{\label{fig:E}The (gravitational) potential energy, kinetic energy, internal potential energy and total energy of the condensate, given in Eq.~(\ref{eq:E}), plotted as a function of the scattering length $a$. In our chosen units, the particle number $N$ is in units of $10^{76}$ particles ($N=2$, i.e., $2\times 10^{76}$ particles, corresponds to one solar mass.) The scattering length is measured in km, whereas the energy is in units of $10^{48}$~J.}
\end{figure}

Our simulation also yielded values for the gravitational potential, internal, and kinetic energies of the condensate. These show a consistent dependence on the scattering length, as shown in Fig.~\ref{fig:E}. As expected, the total energy is negative in all cases, which is an obvious condition for stability.

Lastly, we note that in accordance with Eq.~(\ref{eqn:transformations}), a change of units such that $[{\rm L}]\rightarrow\lambda[{\rm L}]$, $[{\rm T}]\rightarrow\lambda^2[{\rm T}]$, and $[{\rm M}]\rightarrow \lambda^{-1}[{\rm M}]$ (corresponding to unit transformations such that $\lambda^2=\kappa^{-1}$) leaves all results unchanged. Thus, a 1-second simulation of a 1 solar mass Bose star with a 50~km radius, consisting of $5.6\times 10^{-11}$~eV particles is identical to a 100-second simulation of a 0.1 solar mass Bose star with a 500~km radius, consisting of $5.6\times 10^{-12}$~eV particles. Therefore, our numerical results on stability characterize entire classes of BEC objects related to each other by such transformations of units.

\section{Conclusions}

Our simulations demonstrate that a self-gravitating BEC is stable, and that any apparent instability in the simulation is purely an artifact of the finite time and spatial resolution of the simulation itself. By suitably choosing nonphysical parameters, we can achieve both rotating and nonrotating stable Bose stars.

The main limitation of our approach is that we numerically evolve the BEC wavefunction in space, and thus the spatial grid size used in the numerical model must be smaller than the anticipated wavelength of the wavefunction. We estimate this wavelength as the Compton-wavelength of the BEC particle, which yields an upper limit on the BEC particle mass for a given object size and grid resolution. (A possible way to overcome this limitation is to run the simulation in Fourier space \cite{TakPong2009}.)

Our simulation is based on the non-relativistic Gross-Pitaevskii equation, which loses its validity in the relativistic regime. In practical terms, this means a lower limit on the BEC star's radius for a given mass. To explore more compact objects, such as objects with the size of a neutron star, relativistic code will be required.

Presently, our simulation code can only model ``pure'' BEC stars containing little or no normal matter. While this may be good enough to explore the theoretical stability of a self-gravitating BEC object, clearly any real star would contain significant quantities of non-relativistic and relativistic gas. A future simulation may incorporate the (relativistic) hydrodynamic equations that govern the behavior of this gas, as well as any interaction terms between the BEC and the gas.

Even with these limitations, our simulation results are instructive. One particular topic of interest is the apparent lack of vortex formation, which we believe is due to the fact that any vortices forming near the Bose star's surface have a hard time penetrating into the interior. Clearly, this topic of vortex formation in a self-gravitating, rotating Bose star deserves further investigation.

Our most important conclusion remains, however, that a range of possible Bose stars are stable in numerical simulations, and we saw no indication that Bose stars that are beyond our present capabilities (purely due to computational constraints) would behave differently.

\acknowledgments

We thank T. Harko for valuable advice and discussions.

\bibliography{refs}

\end{document}